\documentclass[a4paper,fleqn]{aa}

\usepackage{graphicx}
\usepackage{amsmath}
\usepackage{natbib}
\usepackage{comment}
\usepackage{amssymb}

\begin{document}

\title{Dust-UV offsets in high-redshift galaxies \\ in the Cosmic Dawn III simulation}

\author{Pierre Ocvirk\inst{1},
Joseph S. W. Lewis\inst{2},
Luke Conaboy\inst{3},
Yohan Dubois\inst{2},
Matthieu Bethermin\inst{1},
Jenny G. Sorce\inst{4,5,6},
Dominique Aubert\inst{1},
Paul R. Shapiro\inst{7},
Taha Dawoodbhoy\inst{7},
Joohyun Lee\inst{7},
Romain Teyssier\inst{8},
Gustavo Yepes\inst{9,10},
Stefan Gottlöber\inst{6},
Ilian T. Iliev\inst{11},
Kyungjin Ahn\inst{12},
Hyunbae Park\inst{13},
Mei Palanque\inst{1}}

\institute{Observatoire Astronomique de Strasbourg, Université de Strasbourg, CNRS UMR 7550, 11 rue de l'Université, 67000 Strasbourg, France \and
Institut d'Astrophysique de Paris, UMR 7095, CNRS, UPMC Univ. Paris VI, 98 bis boulevard Arago, 75014 Paris, France \and
School of Physics and Astronomy, The University of Nottingham, University Park, Nottingham, NG7 2RD, UK \and
Univ. Lille, CNRS, Centrale Lille, UMR 9189 CRIStAL, F-59000 Lille, France \and
Universit\'e Paris-Saclay, CNRS, Institut d'Astrophysique Spatiale, 91405, Orsay, France \and
Leibniz-Institut f\"{u}r Astrophysik, An der Sternwarte 16, 14482 Potsdam, Germany \and
Department of Astronomy, The University of Texas at Austin, Austin, TX 78712-1083, USA \and
Institute for Theoretical Physics, University of Zurich, Winterthurerstrasse 190, CH-8057 Zürich, Switzerland \and
Departamento de Física Teórica M-8, Universidad Autónoma de Madrid, Cantoblanco, 28049, Madrid, Spain \and
Centro de Investigación Avanzada en Física Fundamental (CIAFF), Universidad Autónoma de Madrid, 28049 Madrid, Spain \and
Astronomy Center, Department of Physics \& Astronomy, Pevensey II Building, University of Sussex, Falmer, Brighton BN1 9QH, United Kingdom \and
Chosun University, 375 Seosuk-dong, Dong-gu, Gwangjiu 501-759, Korea \and
Lawrence Berkeley National Laboratory, CA 94720-8139, USA \and
Berkeley Center for Cosmological Physics, UC Berkeley, CA 94720, USA \and
Center for Computational Sciences, University of Tsukuba, 1-1-1 Tennodai, Tsukuba, Ibaraki 305-8577, Japan}

\date{Received date / Accepted date}

\abstract{
Recent observations have revealed puzzling spatial disparities between ALMA dust continuum and UV emission as seen by HST and JWST in galaxies at $z=5-7$ (e.g. ALPINE and REBELS surveys), compelling us to propose a physical interpretation of such offsets. 
We investigate these offsets using the Cosmic Dawn III (CoDa III) simulation, a state-of-the-art fully coupled radiation-hydrodynamics cosmological simulation, which incorporates a dynamical dust model. First of all, we find that our simulated dust masses, while calibrated to match observed ones, yield unrealistically large UV attenuations. In fact, the bright-end galaxy UV Luminosity function is best reproduced using only 7.5\% of the dust content of CoDa III galaxies. With this recalibration, we obtain populations of massive galaxies matching ALPINE and REBELS magnitudes and UV slopes, but with smaller dust masses than observed. In this framework, we also find significant dust-UV offsets in massive, UV-bright galaxies ($\mathrm{M}_\mathrm{DM}> 10^{11.5}$ M$_\odot$, M$_*>10^{10}$ M$_\odot$, M$_{\rm AB1600}<-21.5$), reaching up to $\sim 2$ pkpc for the most massive systems.
Our analysis reveals that these offsets primarily result from severe dust extinction in galactic centers rather than a misalignment between dust and stellar mass distributions. At the spatial resolution of CoDa III (1.65 pkpc at z=6), the dust remains in majority well-aligned with the bulk stellar component, and we predict the dust continuum should therefore align well with the stellar rest-frame NIR component, less affected by dust attenuation. This study highlights the importance of dust in shaping the appearance of early galaxies at UV wavelengths, even as early as in the Epoch of Reionization.
}



\titlerunning{dust-UV offsets in Cosmic Dawn III galaxies}
\authorrunning{Ocvirk, P. et al.}

\maketitle

\section{Introduction}

The formation and evolution of galaxies in the early universe, particularly during the epoch of reionization (EoR), remains one of the most intriguing and challenging areas of study in modern astrophysics \citep{Dayal2018}. This era, spanning roughly from redshift $z \sim 20$ to $z \sim 5.3$ \citep{bosman2022}, saw the first stars and galaxies ionize the neutral intergalactic medium (IGM), fundamentally altering the cosmos \citep{barkana_loeb2001,wise2019,Ocvirk2016}. However, our understanding of the physical processes governing galaxy formation during this period is still limited, largely due to observational challenges and the complex interplay of various astrophysical phenomena.

Recent advancements in observational capabilities have begun to shed light on the properties of high-redshift galaxies ($z = 4-7$). The Atacama Large Millimeter/submillimeter Array (ALMA) has been particularly instrumental in this regard, with large programs such as ALPINE (ALMA Large Program to Investigate C$^+$ at Early Times) providing unprecedented insights into the interstellar medium (ISM) and dust properties of galaxies at $z \sim 4-6$ \citep{Bethermin2020, LeFevre2020}. These observations have revealed that a significant fraction of star formation at these redshifts is already obscured by dust, with some galaxies showing substantial far-infrared (FIR) emission \citep{Fudamoto2020, Khusanova2021}.

Building upon these results, subsequent surveys like REBELS (Reionization Era Bright Emission Line Survey) have extended observations to higher redshifts ($z > 6.5$), using ALMA to detect dust continuum and [C II] line emission in UV-selected galaxies \citep{bouwens2022, Inami2022}. These studies have uncovered unexpected features, notably significant spatial offsets between the dust continuum emission detected by ALMA and the ultraviolet (UV) starlight observed by HST and JWST \citep{Bowler2022, Inami2022,killi2024,Dayal2022}. The ALPINE sample, for example, contains about 30\% of objects with offsets ranging from 3.5-4.5 pkpc.
Such offsets challenge our current models of galaxy formation and evolution, suggesting a more complex structure of star formation and dust distribution in early galaxies than previously thought.

The presence of dust in high-redshift galaxies is itself a topic of considerable interest. Dust plays a crucial role in galaxy evolution, affecting the thermal balance of the ISM, catalyzing the formation of molecular hydrogen, and absorbing UV photons to re-emit in the infrared \citep{draine2003}. The detection of substantial dust masses in some galaxies as early as $z \sim 7-8$ raises questions about dust production and its properties, and growth mechanisms in the early universe \citep{Bakx2021,Sommovigo2022,Choban2024,Algera2024}. Moreover, the fraction of obscured star formation appears to increase rapidly with stellar mass and may evolve significantly with redshift \citep{Fudamoto2020}, impacting our understanding of the cosmic star formation history. Crucially, when trying to quantify obsecured star formation density, observers need to rely on accounting for dust-obscured star formation based on  i) the IRX-$\beta$ diagram or ii) correcting for the attenuation measured in the visible or the UV. However, both methods can underestimate the amount of obscured star formation when UV and IR emission are not co-spatial. For more details on these two points, see \cite{Ferrara2022, Sommovigo2020, Sommovigo2022}.

Theoretical efforts to understand these phenomena has driven an effort to include the relevant physics in recent cosmological simulations. The goal of the Cosmic Dawn (CoDa) Simulation Project\footnote{\url{https://coda-simulation.github.io/}} is to model galaxy formation during the EoR, employing fully coupled radiation-hydrodynamics simulations to self-consistently model the interplay between galaxy formation and cosmic reionization \citep{Ocvirk2016, Ocvirk2020}. 

In parallel, other groups have made significant contributions to our theoretical understanding, with, e.g. the Renaissance  \citep{Oshea2015} and the FirstLight simulations \citep{ceverino2017}. The SPHINX simulations \cite{Rosdahl2018} explored the role of binary stars in driving reionization, while Gnedin and collaborators have developed the Cosmic Reionization On Computers (CROC) project, providing insights into the reionization process and the properties of high-redshift galaxies \citep{Gnedin2014, Gnedin2016}. Others have explored the role of AGNs in reionizing the Universe \citep{Trebitsch2021,garaldi_thesan_2022}.
Specific simulation suites also focused on high-z galaxy morphology in the UV and IR, such as \cite{arata2019,pallottini22,punyasheel24}.


These theoretical efforts have highlighted the importance of properly modeling various physical processes, including stellar feedback, metal enrichment, and dust physics.

Hydrodynamical simulations can now incorporate the ab initio evolution of a dust component separately from the metal gas content \citep{bekki2013,mckinnon2016,Li2019,Trebitsch2021,Lewis2023}. Some simulations can follow various grain properties such as their sizes \citep{aoyama2017} or their chemical compositions
\citep{Graziani2020,choban2022} or both \citep{gjergo2018,granato2021,li2021}, which models embedded in a porous multiphase interstellar medium allow to directly account for observed local extinction curves \citep{dubois_dust}.

Building on these advances, 
Cosmic Dawn III, the latest iteration of the Cosmic Dawn Project, incorporates an early version of the \cite{dubois_dust} dust model, allowing for the follow-up of dust formation, distribution, and its effects on observable properties of galaxies as well as, e.g., ionizing escape fractions \citep{Lewis2023}. This improved modeling is crucial for understanding the complex relationship between star formation, dust, and the observed properties of high-redshift galaxies. 

In this paper, we present results from the CoDa III simulation that provide new insights into the origin and nature of the observed dust-UV offsets in high-redshift galaxies. Our simulations show that for massive galaxies (dark matter halo masses of $10^{11} - 10^{12} $M$_\odot$), the central regions, while hosting the highest star formation rates, are also the most dust-obscured. This leads to a scenario where the peak of star formation activity is heavily attenuated and sometimes undetectable in UV, while the observable UV emission comes from less attenuated, but also less intensely star-forming regions offset from the galaxy center. This phenomenon, emerging naturally from our simulations, offers a compelling explanation for the observational findings and highlights the complex interplay between star formation, dust distribution, and the resulting observed morphologies of high-redshift galaxies. The Cosmic Dawn III simulation's large volume, close to ($\sim 100$ Mpc)$^3$, is required in addressing this topic, since ALMA observations at z=4-7 are limited to the brightest, therefore rarest, population of galaxies.
This analysis not only helps explain the observed dust-UV offsets but also contributes to our broader understanding of galaxy formation and evolution during the epoch of reionization.

\section{Methodology}
\subsection{The CoDa III Simulation}

Our study utilizes the Cosmic Dawn III (CoDa III) simulation, a state-of-the-art, fully coupled radiation-hydrodynamics cosmological simulation of galaxy formation during the Epoch of Reionization. CoDa III is the latest iteration in the Cosmic Dawn Project, building upon the successes of its predecessors \citep{Ocvirk2016, Ocvirk2020}. The simulation was performed using the RAMSES-CUDATON code \citep{Lewis2023}, which couples the RAMSES code for N-body dynamics and hydrodynamics \citep{Teyssier2002} with the ATON radiative transfer module \citep{Aubert2008}.

CoDa III simulates a comoving volume of 94.4 cMpc on a side, with a grid resolution of 8192$^3$ cells, providing a spatial resolution of 11.53 comoving kpc, corresponding to a cell size of respectively 1.92, 1.65 and 1.44 physical kpc at redshifts 5, 6, and 7. The simulation follows the evolution of dark matter, gas, stars, and ionizing radiation from redshift z = 150 to z = 4.6. For a comprehensive description of the simulation setup, including cosmological parameters, star formation and feedback prescriptions, and radiative transfer methods, we refer the reader to Lewis et al. \citep{Lewis2023}. See also \cite{sorce2016} and \cite{sorce2018} for the initial conditions generation.

Part of what makes CoDa III  a desirable setup for the present study is the fact that it produces a faithful description of the high-redshift, reionizing universe. Indeed, \cite{Lewis2022} shows that the simulation is in excellent agreement with a number of observational diagnostics and forecasts of the intergalactic medium, namely:
\begin{itemize}
    \item The evolution of the neutral and ionized gas fractions with redshift, related to a non-monotonic cosmic ionizing emissivity as described in \cite{ocvirk2021} and the use of the true speed of light rather than a reduced speed of light approximation, with a uniform or even spatially-varying reduction factor\citep{aubert2019,ocvirk2019}.
    \item The evolution of the ionizing rate with redshift.
    \item the electron-scattering optical depth seen by the cosmic micro-wave background.
    \item The evolution of the ionizing photon mean free path up to z=6 \citep{Lewis2022}.
\end{itemize}
Thanks to these studies, we know that CoDa III's galaxy population reionises the universe in a faithful manner.

\subsection{Dust Model}
\label{s:dustmodel}
The dust model implemented in CoDa III is the same as that included in the  DUSTiER simulation \citep{Lewis2023}. It accounts for dust production, growth, and destruction processes. Dust is produced by supernovae and AGB stars, grows through accretion of gas-phase metals in the interstellar medium, and is destroyed by supernova shocks and thermal sputtering. Dust is accounted for in the ionizing radiative transfer module of CoDa III through its opacity at 611 \AA ${}$ as this is the only photon group we consider here. The dust model assumes a single grain size, and no chemical composition, which limits the ability to predict the shape of the extinction curve (useful to characterize exactly the extinction in the UV). For this reason, we have to assume an extinction curve before running the simulation. CoDa III and DUSTiER chose to employ a Small Magellanic Cloud (SMC) extinction curve\footnote{We remind here the reader that employing instead a Large Magellanic Cloud extinction curve can lead to significantly different results, as shown in the appendix of \cite{Lewis2023}.} from \cite{Draine2001}\footnote{\url{https://www.astro.princeton.edu/~draine/dust/extcurvs/kext_albedo_WD_SMCbar_0}}, as explained in \cite{Lewis2022,Lewis2023}, yielding the following dust absorption coefficients, in m$^2/$kg of dust (not mixture):

\begin{tabular}{lll}
    $\kappa_{\rm{d},1500 \, \text{\AA}}$ & = & $4.89 \times 10^3$ \\ 
    $\kappa_{\rm{d},1600 \, \text{\AA}}$ & = & $3.83 \times 10^3$ \\
    $\kappa_{\rm{d},2500 \, \text{\AA}}$ & = & $8.82 \times 10^2$
\end{tabular}

The dynamical dust model parameters (dust condensation efficiency, the maximum dust-to-metal ratio, characteristic timescale for dust growth ) were calibrated to reproduce  observed dust masses in high-redshift galaxies by various studies \citep{Mancini2015,Burgarella2020}, including those detected by the REBELS survey \citep{Dayal2022}. For a detailed description of the dust model and its calibration, we refer the reader to \cite{Lewis2022,Lewis2023}.
By incorporating this calibrated dust model, CoDa III aims to provide a simple representation of dust content and distribution in high-redshift galaxies, allowing us to investigate the relationship between dust emission and UV starlight and its impact on observable properties of the galaxies.
However, as we will explain in more detail in Sec. \ref{s:dustrecal}, using the raw dust masses from the simulation yields unrealistically high dust attenuations, which prompted us to multiply all dust masses by a K$_{\rm dust}=0.075$ factor in the rest of the paper.

\subsection{Galaxy Identification and Definition}
\label{s:fof}
To identify and define galaxies within our simulation, we
employ a friends-of-friends (FOF) algorithm implemented
in the Parallel Friends-of-Friends (PFOF) code \citep{roy-pfof}. The choice of linking length is crucial in FOF algorithms,
as it determines the extent to which particles are
grouped together into a single structure. The linking length determines the density at the boundary of the FOF object, e.g. 81.62 in case of a linking length 0.2 \citep{more2011}.
  However the resulting overdensity of the object depends also on the concentration which depends on redshift. Thus the same linking length leads to different overdensities at a given redshift, and the mean overdensity decreases for a given linking length at higher redshifts due to concentration evolution. 

Initially, we experimented at high redshifts with a linking length of 0.2 
times the mean inter-particle separation, which is a common
choice in cosmological simulations at z=0. However, we found
that this led to overlinking, partly due to an artificial
merging of distinct galactic structures. To address this
issue, we checked smaller linking lengths and adopted a value of 0.15
times the mean inter-particle separation. This choice strikes
a balance between identifying coherent galactic structures
and avoiding spurious linking between separate galaxies. Moreover, at high redshifts the FOF mass function is in good agreement with a spherical overdensity  mass function which assumes a spherical overdensity of 200 \citep{watson2013}. This confirms that 0.15 is an appropriate choice.

Using the FOF masses of the structures identified with this linking length, we then compute for each halo a characteristic radius R$_{\rm FoF}$ using an overdensity of 200, which we use in the following analysis to identify the stellar particles residing in the given halo. This approach is consistent with the methodology used
in the DUSTiER simulation \citep{Lewis2023}, although
the latter used the ramses built-in halo finder PHEW \citep{PHEW}.
By defining galaxies in this manner, we ensure a robust
identification of galactic structures while maintaining consistency
with established methods in the literature. This
definition forms the basis for our subsequent analysis of the
spatial distribution of dust and UV emission within high-redshift
Galaxies.

\subsection{High-z galaxies in CoDa III: an illustrative overview}
\label{s:illustration}

\begin{figure*}[!t]

\includegraphics[trim=30mm 10mm 30mm 10mm,clip,width=1.\linewidth]{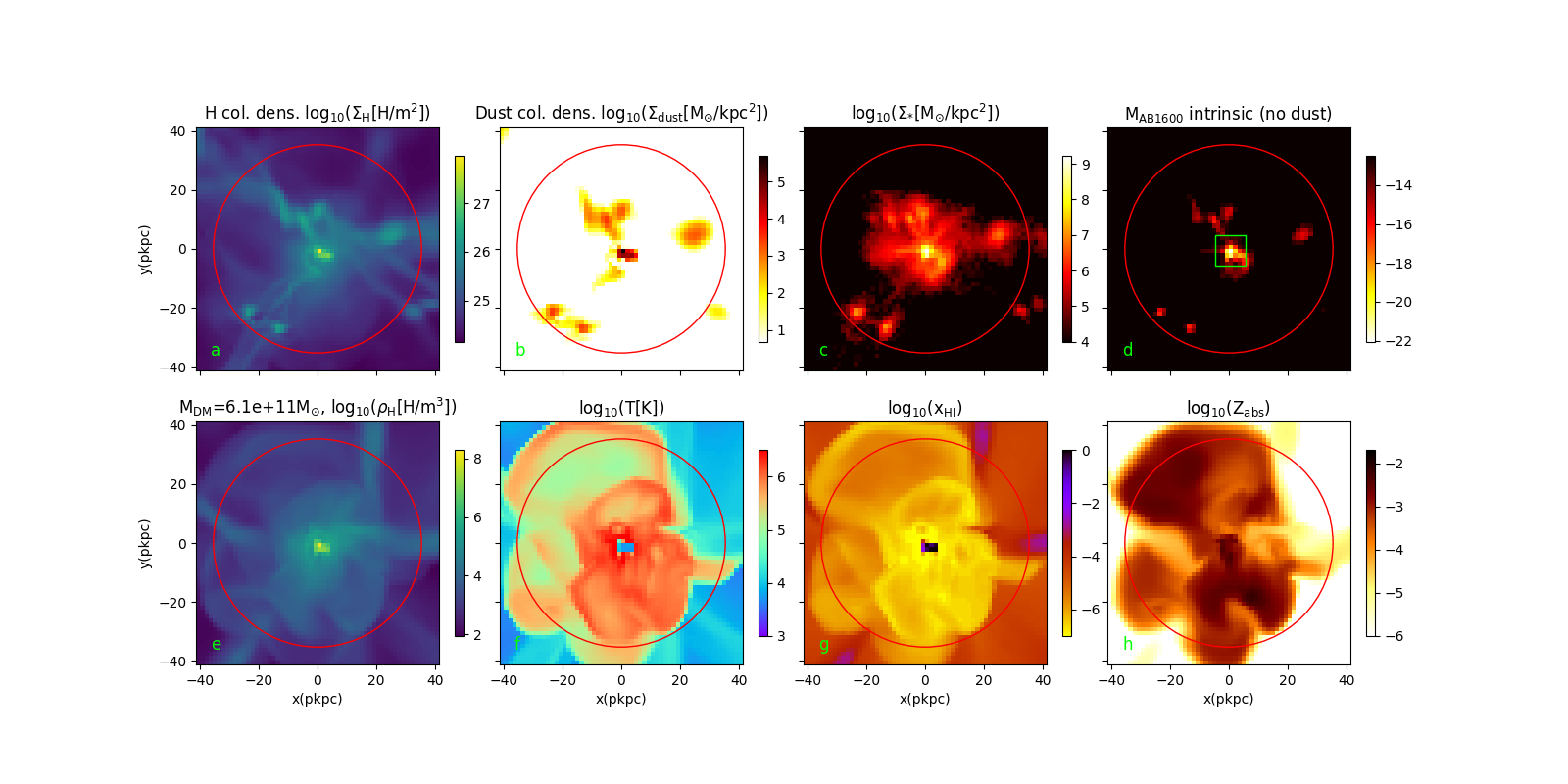}
\caption{Physical properties of a representative galaxy at z=7 with M$_{\rm DM} = 6.1 \times 10^{11}$ M$_{\odot}$. The red circle in each panel indicates R$_{\rm FoF}$. Top row: Projected quantities integrated along the line of sight. Bottom row: 1-cell thick slice at the position of maximum density. (a) H column density, (b) Dust column density, (c) Stellar mass surface density, (d) Intrinsic (unattenuated) absolute UV magnitude at 1600 \AA $ $, (e) H number density, (f) Temperature, (g) HI fraction, (h) Metallicity (absolute scale).}
\label{fig:galaxy_properties}

\end{figure*}

To illustrate the appearance of galaxies in CoDa III, we show in Fig. \ref{fig:galaxy_properties} maps of several properties of the baryons in a massive M$_{\rm DM}$=6.1x10$^{11}$ M$_{\odot}$ z=7 galaxy. All coordinates and quantities use the physical reference frame as opposed to co-moving, hence the 'pkpc' in the labels. The red circle in each panel denotes R$_{\rm FoF}$, the friends-of-friends radius of the halo as defined in Sec. \ref{s:fof}

The top row shows projected quantities integrated along the line of sight:
\begin{itemize}
    \item Panel (a) displays the H column density, revealing the overall gas distribution of the galaxy halo and its surroundings. We describe the gas distribution from the center outwards. The gas distribution features a central concentration, with some elongation, i.e. it is not spherical. It also shows signs of perturbations, such as a gaseous arc to the bottom left of the center, reminiscent of a gas tidal stream, suggesting ongoing or very recent merging activity, contributing to the disturbed appearance of the central structure. 
    An infalling group of gas structures is also seen above and to the left of the halo center, supporting the notion that merging is intense and frequent in this halo and at this epoch. Further out, filaments extend out all the way to and beyond R$_{\rm FoF}$. Along these filaments, smaller gas clumps can be found, corresponding to a population of smaller, infalling sub-haloes and galaxies.
    \item Panel (b) shows the dust column density, which is also concentrated in the central regions, with smaller clumps co-located with the infalling sub-haloes of panel (a). The 0.075 factor justified in Sec. \ref{s:dustrecal} is accounted for.
    \item Panel (c) presents the stellar mass surface density. It features a centrally concentrated stellar component, with a few clumps corresponding to the infalling sub-haloes. It also displays some extended, low density structure.
    \item Panel (d) depicts the intrinsic (unextincted) absolute UV magnitude at 1600 \AA $ $, highlighting the regions of recent star formation. Indeed, only the latter have a significant output at 1600 \AA $ $. The extended structure of panel (c) is less prominent in this panel, because it corresponds to an older, passively evolving stellar component.
\end{itemize}

The bottom row presents 1-cell thick slices at the position of maximum density, allowing us to see fine structures that are not visible in the projected maps:
\begin{itemize}
    \item Panel (e) shows the H number density slice, revealing detailed gas structures including filaments and clumps, but also gas shells typical of SN winds.
    \item Panel (f) displays the temperature slice, where the high temperature of the gas shells seen in (e) confirms their origin as SN-induced shocks and winds, reaching all the way to $R_{\text{FoF}}$ and in places slightly beyond. The several quasi-concentric shells likely originate from a series of distinct, strong episodes of star formation. Remarkably, the most central region of the galaxy remains cool despite the intense star formation activity. Beyond $x_{\text{HI}} = 10^{-2}$, the un-shocked intergalactic medium sits at a typical photo-heated temperature of $\sim 10^4$ K.
    \item Panel (g) presents the neutral Hydrogen fraction slice. showing the ionization state of the gas. The halo and its surroundings are highly ionized except for the very center of the galaxy, which remains neutral. Some fractions of connecting filaments, at or beyond R$_{\text{FoF}}$, are moderately neutral, with x$_{\rm HI}$=$10^{-2}$. The rest of the halo is strongly ionized, reflecting the high-temperatures of the circum-galactic medium, with extreme ionizations shells (x$_{\rm HI}$=$10^{-6}$) accompanying the SN shocks. Beyond the range of SN winds, the photo-ionized intergalactic medium has has a neutral fraction slightly below x$_{\rm HI}$=$10^{-4}$.
    \item Panel (h) shows the metallicity slice, illustrating the enrichment of the ISM by stellar feedback, with higher metallicities in the galaxy center and in outflowing gas, and uncontaminated, pristine gas beyond R$_{\rm FoF}$.
\end{itemize}

This multi-panel view provides a comprehensive picture of the galaxy's physical properties, showcasing the complex interplay between gas, dust, stars, and feedback processes in high-redshift galaxies. Altogether, the structure we find is typical of simulations of early galaxies, their CGM and surrounding IGM, and is similar to the distributions seen in e.g. \cite{ocvirk2008} and \cite{Ocvirk2016}. We note however generally more prominent cold neutral cores in massive galaxies than portrayed in  \cite{Ocvirk2016}'s figures. This is likely a consequence of the increased mass and spatial resolution compared to CoDa II and of the changes in the sub-grid star formation model.

The dust-UV offsets we propose to investigate take place at very small scales, in the center-most regions of our galaxies. To help the reader to grasp the different scales of the dataset, we show in as a green square in panel (d) the size and position of the regions we will consider and analyse in our galaxy sample through the rest of the paper.

\subsection{Attenuated UV Map Generation}
\label{s:attuvmapgen}

To produce synthetic observations for each galaxy, we extract the dust data cube from the simulation for the region containing a dark matter halo and its associated stellar particles. We center this cube on the peak of gas density within the halo. The intrinsic UV emission from stellar particles is calculated using the Binary Population and Spectral Synthesis (BPASS) models \cite{Eldridge2017}, consistent with the approach used in the CoDa III and DUSTiER simulations \cite{Lewis2022, Lewis2023}.

To simulate the absorption of UV photons by dust, we perform a one-dimensional radiative transfer calculation along the z-direction, treating this direction as the line of sight to the observer. The dust grain properties, including the extinction curve, are based on the SMC model by \cite{Draine2001}, as detailed in the DUSTiER paper \cite{Lewis2023} and Sec. \ref{s:dustmodel}. This ensures consistency in our treatment of dust across different aspects of the simulation and analysis\footnote{The corresponding extinction curve, along with other models, can be found here \url{https://www.astro.princeton.edu/~draine/dust/dustmix.html}}. 

{\bf{For simplicity, we account exclusively for absorption and do not consider the albedo of the grains and subsequent reflexion/diffusion of incident radiation. As a consequence, it may appear that our treatment may overestimate the dust attenuation. However, since the absorption cross-section we use is only (1-albedo) times the interaction cross-section, our treatment is similar to the limit case of a model with reflection on dust grains, but where the reflected radiation (the albedo) is emitted in the same direction as the incident radiation. Because of the isotropic nature of stellar sources, and the roughly spherical nature of our gas and dust distributions at the center of CoDa III galaxies, we expect a full accounting of reflexions and diffusions would not significantly alter our results, although this is admittedly a limitation of our study that could be probed with a more accurate dust radiative transfer post-processing such as SKIRT\footnote{\url{https://skirt.ugent.be/root/_home.html}} \citep{skirt} or RASCAS\footnote{\url{http://rascas.git-cral-pages.univ-lyon1.fr/rascas/}} \citep{rascas}.}}



\subsection{Dust mass recalibration}
\label{s:dustrecal}
The impact of dust on the UV luminosity function is strongest at the bright-end. Due to its limited size (16 h$^{-1}$ cMpc), DUSTiER \citep{Lewis2023} could not probe this aspect beyond M$_{\rm AB1600}<-21$. When analysing Cosmic Dawn III, which ran using the DUSTiER calibration of the dust model, we found that in order to reproduce the UV galaxy luminosity function's bright end, we had to reduce the raw simulation dust masses by a factor K$_{\rm dust}$=0.075, and all the dust masses and densities used in the present paper take this factor into account. We note however here that such a recalibration of the dust masses should not affect the reionization history obtained for CoDa III. Indeed, as found in \cite{Lewis2023}, the galaxy ionizing escape fractions are dominated by HI opacity and the dust $<912$ \AA $ $ opacity is sub-dominant at all halo masses and all redshifts of the simulation.

\begin{figure}[t]
\includegraphics[width=1.05\linewidth]{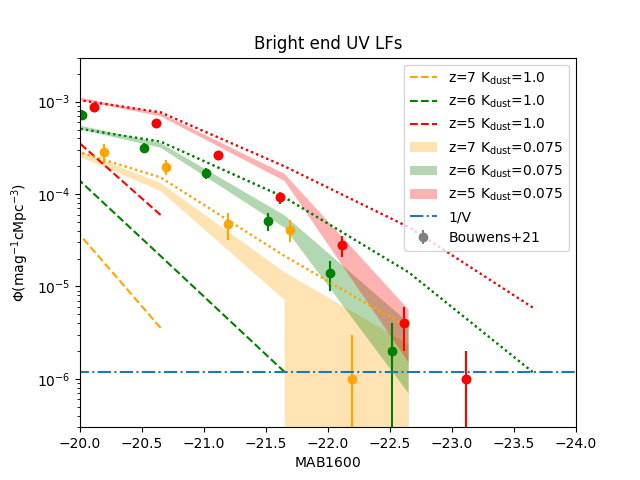}
\caption{{\bf Bright end UV luminosity function of CoDa III galaxies for redshifts z=5,6,7 and various dust contents. The shaded areas represent the average dust-attenuated LFs plus or minus the Poisson noise for K$_{\rm dust}$=0.075. The dotted lines represent the unattenuated LFs. The dashed lines represent the full dust content LFs (i.e. K$_{\rm dust}$=1.0). The horizontal dot-dashed line represents 1/V limit where V is the volume of the simulated domain in Mpc$^3$. The observed LFs are from \cite{bouwens21LF}.}}
\label{f:uvlf}
\end{figure}

{\bf To compute the UVLFs presented in Fig. \ref{f:uvlf} we proceed as follows. For any galaxy, we generate its attenuated UV map and integrate its flux spatially. Our LFs therefore account for dust attenuation by construction. The LFs are shown for several dust contents: the fiducial K$_{\rm dust}$=0.075 reduction factor (shaded areas), the full dust content (dashed colored lines), and no dust (dotted colored lines). The full dust content K${\rm dust}$=1.0 LFs are incompatible with the observations. Reducing the dust content moves the LFs towards the unattenuated limit. We settled on the K$_{\rm dust}$=0.075 value as it yielded the best agreement with the observed LFs of \cite{bouwens21LF}. This trend has also been reported in the literature. In \cite{Dayal2022}, for instance, the model that best fits the dust masses undershoots the UVLFs, whereas the model that best reproduces the UVLFs has a  smaller dust content. More recently, difficulties to match high-z LFs using the very high dust masses suggested by ALPINE/REBELS have also been reported in \cite{toyouchi25-dust-lf}, \cite{zhao24}.}

\begin{figure}[th]
    \centering
    \includegraphics[width=0.95\linewidth]{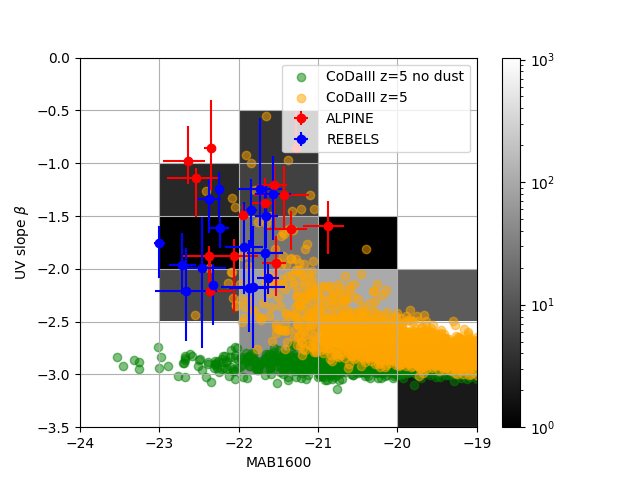}
    \includegraphics[width=0.95\linewidth]{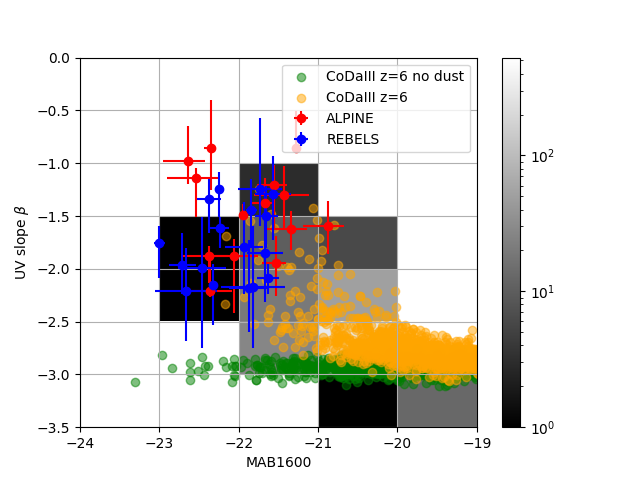}
    \includegraphics[width=0.95\linewidth]{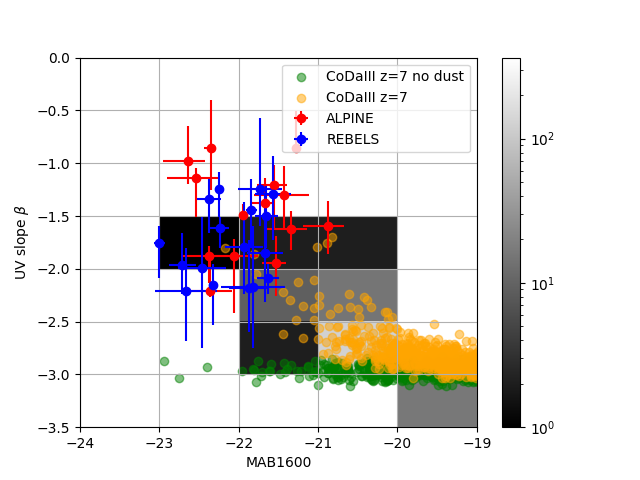}
    \caption{M$_{\rm AB1500}$-$\beta$ distributions of CoDaIII galaxies (orange: dust-attenuated, green: unattenuated), along with the ALPINE and REBELS samples (respectively red and blue symbols with error bars). The gray-shaded colors indicate how many galaxies (dust-attenuated) reside in each cell, as a visual help. Note that  nebular continuum emission is not accounted for, and can redden the spectra by up to $\Delta \beta=+0.3$ \citep{wilkins2016}.}
    \label{fig:beta}
\end{figure}

We also find support for this choice of our fiducial dust mass recalibration K$_{\rm dust}$=0.075 in the UV slope $\beta$ of our simulated galaxies. To obtain the UV slope we compute 2 attenuated flux maps as explained in \ref{s:attuvmapgen}, in a 200 \AA\,  wide window centered on 1500 and 2500 \AA. We then spatially integrate the map to obtain the fluxes F$_{1500}$ and F$_{2500}$ and compute the UV slope as $\beta=\log(\rm{F_{2500}}/\rm{F_{1500}})/\log(2500/1500)$. The slopes are shown in Fig. \ref{fig:beta} for 3 redshifts (z=5,6,7), along with the ALPINE and REBELS samples detailed in Sec. \ref{s:obs}. For each redshift we show the attenuated and unattenuated (M$_{\rm AB1600}$ , $\beta$) distributions of simulated galaxies.
The impact of dust on the simulated galaxies clearly manifests as a significant dimming and reddening. The brightest, dust-attenuated CoDa III galaxies have a fair amount of overlap with the ALPINE and REBELS samples at z=5. Our results also suggest that galaxies fainter than the REBELS and ALPINE samples should have bluer colors. 
The reasonable overlap between CoDaIII and observed galaxies confirms our fiducial dust mass recalibration K$_{\rm dust}$=0.075.


\begin{figure}
\includegraphics[width=1.05\linewidth,clip=false, trim= 0 -1cm 0 0]{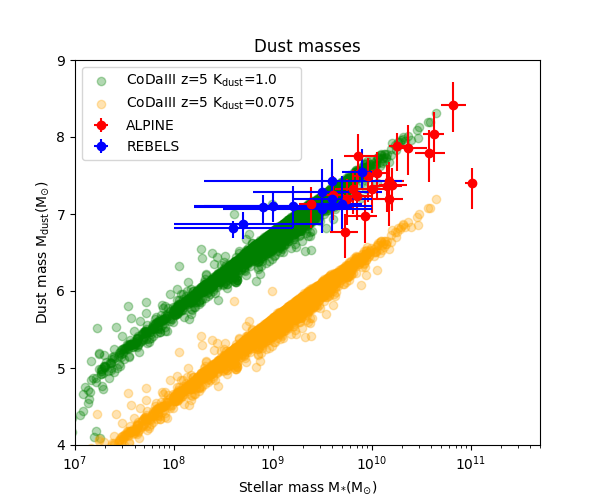}

\caption{{\bf Dust masses in the Cosmic Dawn III simulation at z=5 as a function of stellar mass, along with the ALPINE and REBELS samples. Green and orange colors correspond to 100\% and 7.5\% of the simulated galaxies' dust masses, respectively. Our fiducial model for the post-processing of this study is the 7.5\% case.}}
\label{f:dustmass}
\end{figure}

However, this recalibration degrades the agreement between our simulated dust masses and the observed dust masses, as seen in Fig.  \ref{f:dustmass}. The DUSTiER simulation \citep{Lewis2023} was initially calibrated to match them, and did so successfully, but the reduction factor we introduce to match the UVLFs means we now undershoot the observed dust masses, even including their significant error bars.
We were unable to obtain at the same time a good agreement with the UVLFs at z=5,6,7 {\em and} the dust masses, and had to make a choice, even if this is not fully satisfactory. We decided to prioritize the agreement with the UVLFs, as these may be more secure, since they involve less assumptions in their observational determinations, although that may reflect a personal bias of the authors.

Admittedly, the spatial resolution of CoDa III is limited to >1 pkpc at z=7, which may not fully capture the fine structure of dust and star formation in galaxies. 
In fact, it is reasonable to expect some amount of structure beyond CoDa III resolution. Galaxy effective radii, for instance, have been determined to be often smaller than 1 pkpc at z$\sim 7$, albeit for galaxies fainter than the REBELS sample \citep{yang2022_size_L}, and the galaxy itself may display structure beyond this limit, as shown in e.g. \cite{arata2019} and \cite{pallottini22}.
Despite these limitations, we will see that our approach allows us to make meaningful comparisons between the large-scale distributions of dust and UV emission in our simulated galaxies, which is the primary focus of this study.




\subsection{Dust Maps}

To complement our attenuated UV maps, we also generate dust maps for each galaxy. These maps are created by integrating the three-dimensional dust density distribution along the same z-axis used for the UV maps.

It is worth noting that an accurate estimation of dust continuum emission typically requires accounting for the amount of UV light absorbed by the dust and computing the dust temperature which controls the blackbody mean frequency, as this absorbed energy is re-radiated in the infrared. For the sake of simplicity in this study, we have not included this step in our dust map generation. Instead, we use the dust density projection as a proxy for dust emission. By doing this we assume far-IR diffusion on dust grains does not affect significantly the geometry (and at least the position of the maximum of emission) and the dust temperature is constant across our galaxies.

Since the ALMA beam is comparable to our cell size at z=5-7, we do not degrade our maps through e.g. a spatial convolution, as can be done for higher resolution simulations.

Despite this simplification, our results demonstrate that this approach still effectively captures the key geometrical differences between the dust emission and UV-emitting stellar distributions that are central to this study. While the absolute brightness of our dust continuum maps may not be precisely calibrated, their morphology and relative brightness distribution provide valuable insights into the dust structure of these early galaxies.

\subsection{Observed offsets: ALPINE and REBELS}
\label{s:obs}
Our comparison with observations draws upon data from two major surveys of high-redshift galaxies: ALPINE and REBELS. These surveys provide crucial observational constraints on the spatial relationship between dust and UV emission in early galaxies.

The ALPINE survey \citep{LeFevre2020, Bethermin2020} targeted 118 star-forming galaxies in the redshift range $4.4 < z < 5.9$. For our analysis, we specifically used the subset of ALPINE galaxies that had both a [C II] line detection (providing a spectroscopic redshift) and a measured continuum-UV offset. This selection ensures that we can accurately convert the observed angular offsets to physical distances. It yields 14 galaxies.

The REBELS survey \citep{bouwens2022,Inami2022,bowler2024} focused on 40 UV-bright galaxies at $z > 6.5$. These galaxies were selected based on their photometric redshifts, with spectroscopic confirmation coming from [C II] or [O III] line detections. We retain the galaxies for which offsets, stellar masses and uv slopes $\beta$ were computed in these articles. It is worth noting that we excluded one object from the REBELS sample in our analysis. The galaxy REBELS-19 showed an exceptionally large offset $\sim$8pkpc that appeared to be an outlier compared to the rest of the sample. While this object merits further investigation, we chose to omit it from our current analysis to avoid potentially skewing our results. This leaves us with 17 REBELS galaxies. 

In both ALPINE and REBELS survey, it should be recalled that comparing positions measured with such radically instruments as ALMA and HST/JWST is not trivial. Great care must be taken to establish the error budget of each observation and assess the significance of the offsets. We refer the reader to the discussion in \cite{Bowler2022} on this topic.

To ensure consistency in our comparison, we converted all angular offsets from both surveys to physical distances in kiloparsecs. This conversion was performed using the standard Planck cosmology \citep{Planck2018}, which is the same cosmology used in our Cosmic Dawn III simulations. Specifically, we used the cosmological angular size distance appropriate for each galaxy's redshift.

By combining data from these two surveys, we have assembled an observational dataset of dust-UV offsets in the early universe. This dataset provides a valuable benchmark against which we can compare our simulation results, allowing us to test the validity of our dust models and improve our understanding of the complex interplay between dust and star formation in high-redshift galaxies. 

\begin{figure*}[!t]
\includegraphics[trim=30mm 40mm 30mm 50mm,clip,width=1.\linewidth]{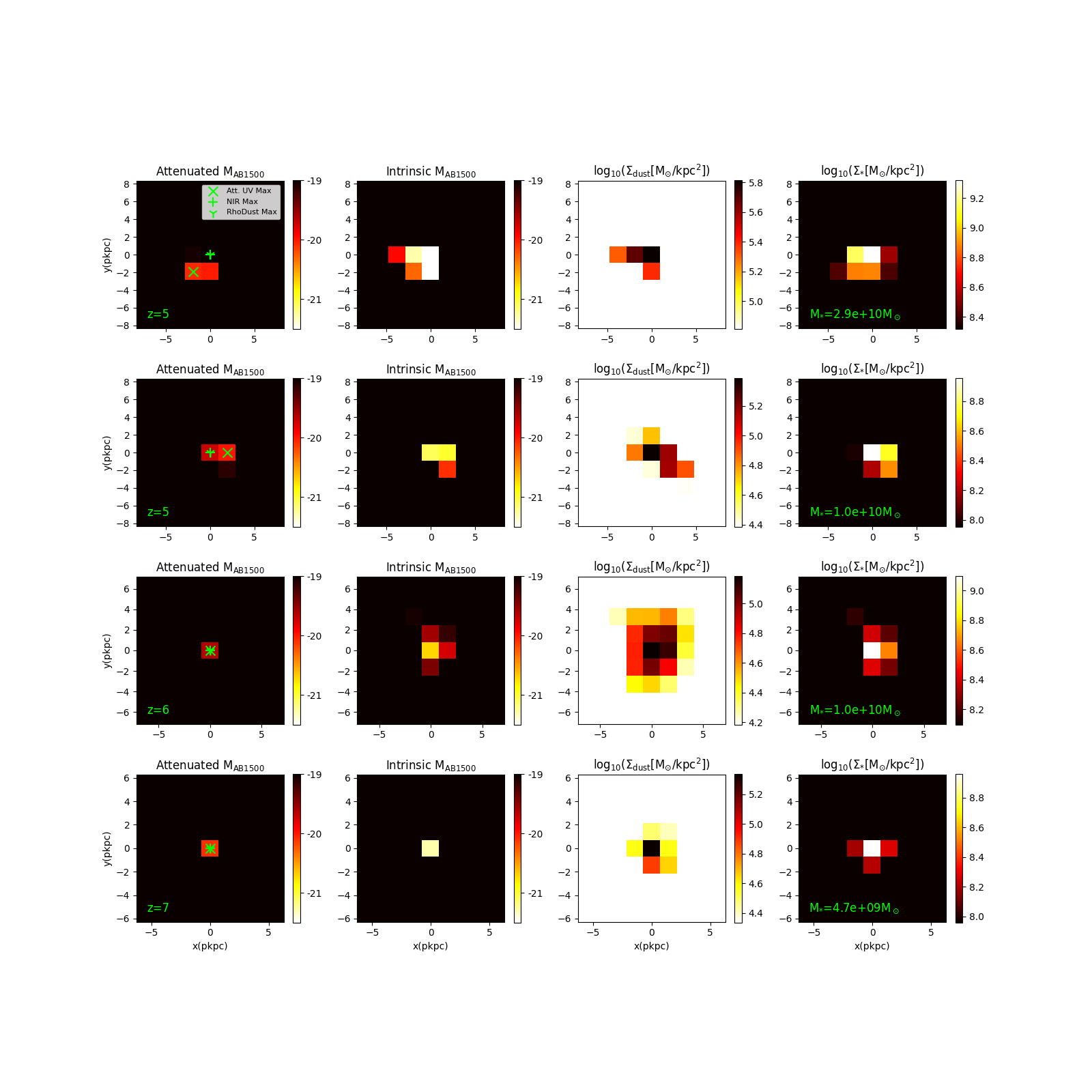}
\caption{From left to right: restframe-UV absolute M$_{\mathrm{AB1600}}$ magnitudes, attenuated and intrinsic (unattenuated), dust column density and projected stellar density maps of a sample of 4 Cosmic Dawn III galaxies with dark matter halo masses M$_{\mathrm{DM}}$=$5\times10^{11}-10^{12}$ M$_{\odot}$ at z=5-7. In the left panel, for each galaxy, symbols mark the positions of maximum attenuated UV flux (green 'x'), maximum NIR flux, using as proxy the maximum of the stellar projected density (green '+'), and maximum dust column density (green 'Y'). For the 2 leftmost plot columns, the color indicates the absolute magnitude (attenuated or intrinsic) of each cell. The dust column density and projected stellar density are given in log$_{10}$(M$_{\odot}$/pkpc$^2$).}


\label{fig:uv_dust_maps}
\end{figure*}


\section{Results}
\subsection{UV and dust maps}

Figure \ref{fig:uv_dust_maps} shows (from left to right) attenuated and intrinsic restframe UV maps, as well as dust column density and projected stellar density for 4 representative massive galaxies M$_{\rm DM}>$ 5x10$^{11}$ M$_{\odot}$ between z=5 and z=7. 
While Fig. \ref{fig:galaxy_properties} intended to illustrate general galaxy structure and used a wide dynamic range, we chose in Fig. \ref{fig:uv_dust_maps} to use a more narrow dynamic range, to be more representative of current instrumental detection limits. This can be easily seen by comparing the color bars of the 2 figures for the same field.

Similar to what we noted in the global halo view in Sec. \ref{s:illustration}, the central distributions of dust and stellar mass are not spherically symmetric. Instead, they display perturbed morphologies, suggestive of ongoing interaction with recently or currently infalling objects and/or merging.




We have also marked the positions of maximum dust column density and maximum attenuated stellar UV flux with symbols. We use the position of the maximum of the stellar density as a proxy for the stellar resframe NIR maximum flux position, since stellar NIR flux is much less attenuated by dust than UV. These maps are projections of the simulation grid, centered on galaxy centers, at the native grid resolution of the simulation. The blocky appearence of the maps therefore reflects the cell size, which is 11.53 ckpc, i.e. 1.65 pkpc at z=6.

We start with a few simple remarks to check the correct behaviour of these outputs. The attenuated maps are significantly fainter than the intrinsic mags maps, as expected due to dust attenuation. The intrinsic UV maps have similar morphology to the dust and stellar density maps, for each galaxy. For all of them, the dust and NIR maximum flux coincide.
However, and this is the most striking aspect of this gallery of maps, the attenuated maps differ significantly from all the others: only a fraction of the stellar distribution is seen in the attenuated maps. More critically, the attenuated UV maximum and the dust maximum, shown by the '+' and 'Y' symbols, are frequently offset (e.g. rows 1 and 2 from the top). This is obviously due to dust extinction, as the faintest cells in the attenuated maps are those with the highest dust column density. Although such dust-UV offsets are common, in particular at the highest halo masses, they are not systematic. The 2 bottom rows, for instance, show  galaxies where all 4 maxima are co-located in the central cell. This configuration is more typical of CoDa III lower mass galaxies.

Such offsets driven by dust attenuation and complex morphology/clumpiness or multiplicity have also been reported in the literature, e.g. \cite{arata2019}.
It has also been noted in \cite{pallottini22} that "...the multi-phase ISM structure in these systems, consisting of IR-emitting molecular clumps embedded in a semi-transparent, UV-emitting diffuse component. This configuration also produces a UV vs. dust continuum spatial offset."
\cite{cochrane2019} reports that UV and far-IR emission occupy strikingly different locations in their simulated galaxies, also driving significant offsets between UV and dust emission. This can also seen in Fig. 4 of \cite{behrens2018}. 
Finally, while \cite{punyasheel2025} does not specifically quantify the uv-dust offsets in their study, their Fig. 5 shows a mixture of clear and insignificant/null offsets.

\subsection{Comparison to observed dust-UV offsets}
\label{s:dust-uv}
\begin{figure}[!t]
\centering
\includegraphics[width=0.5\textwidth]{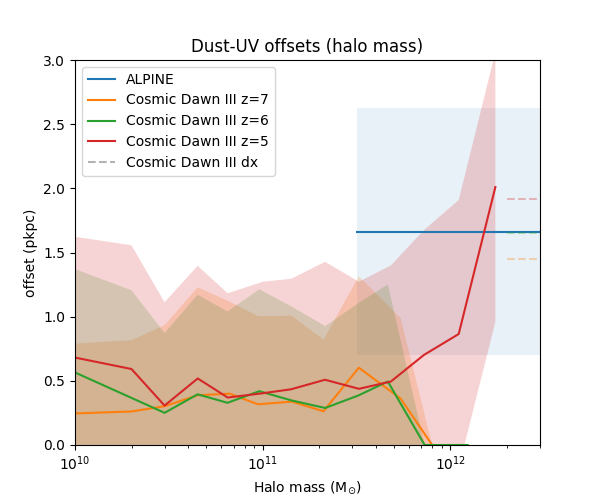}
\caption{Dust-UV offsets as a function of halo mass. The blue shaded region represents the mean and dispersion of observed offsets in the ALPINE survey, with a tentative, rough dark matter halo mass range estimate is used to locate the sample on the horizontal axis. Colored lines show the mean offsets in Cosmic Dawn III simulations at different redshifts (z=7, 6, and 5). Shaded regions around these lines indicate the 1$\sigma$ dispersion. The dashed segments to the right represents the cell size of the simulation.}
\label{fig:dust_uv_offsets}
\end{figure}

To contextualize our findings within the broader observational landscape, we compare the dust-UV offsets found in our Cosmic Dawn III simulations with those observed in high-redshift galaxies. Figure \ref{fig:dust_uv_offsets} presents this comparison.

In Fig. \ref{fig:dust_uv_offsets}, we plot the dust-UV offsets against halo mass. The blue shaded region represents the mean and dispersion of observed offsets from the combined ALPINE and REBELS surveys, providing a benchmark for comparison with our simulations. The colored lines show the mean offsets found in our Cosmic Dawn III simulation at three different redshifts: z=7 (orange), z=6 (green), and z=5 (red). The shaded regions around these lines indicate the 1$\sigma$ dispersion of the offsets in our simulated galaxies. To place the observed sample in this plot we tentatively assign an approximate dark matter halo mass M$_{\rm DM}$=10$^{11.5-12.5}$ M$_{\odot}$ to the ALPINE+REBELS sample. Several  trends emerge from this comparison:

1. The mean simulated offsets increase with dark matter halo mass at z=5, particularly for masses above $10^{11.5}$ M$_{\odot}$. 

2. For the most massive dark matter halos ($\sim  10^{12}$ M$_{\odot}$), our simulated offsets at z=5 are consistent with the observed mean offsets.

3. At lower halo masses, the simulated offsets fall below the cell size of the simulation, making these predictions less reliable. To help the reader gauge this aspect, the cell size is shown by the 3 dashed segments on the right of the figure, representing the physical cell size at each redshift considered. The cell size of the simulation is constant in comoving kpc, and therefore increases with the scale factor. The low mass sample is compatible with a 0 average offset given the significant dispersion we measure.


4. In the light of the dispersion of our measurements for each mass bin, and in particular at low mass where the offsets are smaller than the simulation's spatial resolution, our results are compatible with a scenario where the offset distribution at fixed halo mass does not evolve with redshift between z=5-7, and the lack of large offsets at z>5 arises simply from the lack of massive enough haloes.


\subsection{Dust-NIR distributions}
\label{s:dust-ir}

To further investigate the nature of the dust-UV offsets observed in our simulations, we examine the spatial relationship between dust and the bulk stellar component, as it would be traced by NIR emission. In practice, we just compute the maximum of the projected, unattenuated stellar density, and use it as a proxy for the location of the stellar rest-frame NIR emission.

\begin{figure}[!t]
\centering
\includegraphics[width=0.5\textwidth]{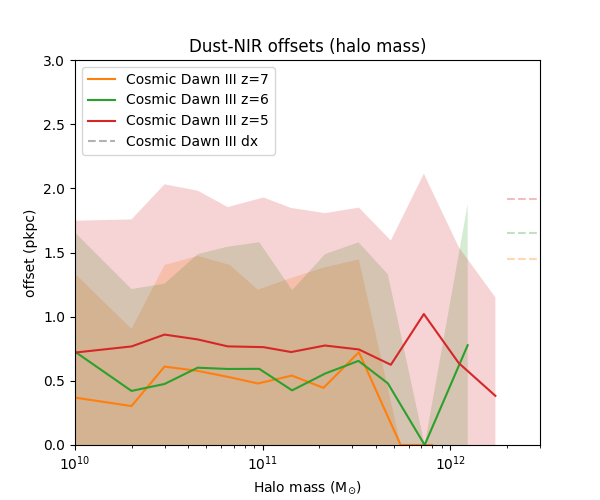}
\caption{Dust-NIR offsets as a function of dark matter halo mass in the Cosmic Dawn III simulations. Colored lines show the mean offsets at different redshifts (z=7, 6, and 5), with shaded regions indicating the 1$\sigma$ dispersion. The dashed segments to the right represent the evolving physical cell size of the simulation (Cosmic Dawn III dx).}
\label{fig:dust_ir_offsets}
\end{figure}

Figure \ref{fig:dust_ir_offsets} shows the offsets between dust and NIR emission as a function of dark matter halo mass for different redshifts.

The key finding here is that the mean dust-NIR offsets do not evolve with dark matter halo mass, unlike the dust-UV offsets as we saw in Sec. \ref{s:dust-uv}.
Their mean value remains well below the simulation's cell size, and, crucially, much smaller than the dust-UV offsets above $10^{11.5}$M$_{\odot}$. The stark contrast between the large dust-UV offsets (Figure \ref{fig:dust_uv_offsets}) and the small dust-NIR offsets suggests that the former are not primarily due to a misalignment between dust and stars. Instead, the dust-UV offsets are principally the result of massive extinction by dust, which obscures the UV emission from the central regions of galaxies where both dust and stars are concentrated.



\subsection{Dust-UV offsets as a function of UV magnitude}

To further facilitate a more direct comparison with observations, we present the dust-UV offsets as a function of the observed (therefore dust-extincted) UV magnitude at 1600 Å (M$_{\rm AB1600}$), in Fig. \ref{fig:dust_uv_offsets_mag}. Several key features appear:

\begin{figure}[!t]
\centering
\includegraphics[width=0.5\textwidth]{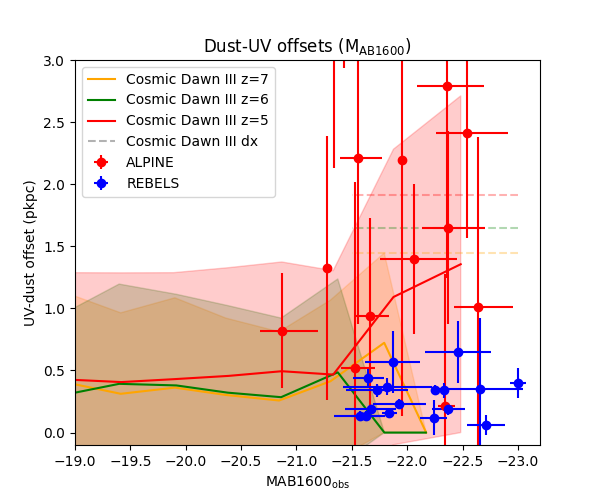}
\caption{As Fig. 3 but using the observed (dust-attenuated) absolute UV magnitude (M$_{\rm AB1600}$).}
\label{fig:dust_uv_offsets_mag}
\end{figure}


1. Trend with UV brightness: The dust-UV offsets generally increase as galaxies become brighter. This trend is consistent across all simulated redshifts and aligns well with the observational data from ALPINE, especially for CoDa III at z=5.

2. The predicted offsets for galaxies brighter than -22 are comparable to the simulation's resolution as per the cell size. This is also the region where the agreement with ALPINE observations is found.

3. The ALPINE offsets are significantly larger than REBELS's despite similar magnitudes. However, the large dispersion of CoDa III offsets at these magnitudes encompasses both ALPINE and REBELS samples.

4. For fainter galaxies (M$_{\rm AB1600}$ > -21), the mean of the simulated offsets falls below the cell size of the simulation (horizontal dashed lines), making these measurements less reliable. 

4. Observational bias: The observational data are limited to relatively bright galaxies (M$_{\rm AB1600}$ < -21). Despite the insufficient spatial resolution in the fainter $>-21$ regime, our results suggest that, taken at face value, dust-UV offsets should be smaller for fainter galaxies, a prediction that will be tested by deeper surveys and future numerical simulations.

5. Redshift evolution: as in the previous figures, there is no clear trend with redshift, beyond the fact that the latest epochs feature naturally more massive haloes. However, at fixed mass, the curves for each redshift fall well within  1 $\sigma$ of each other. Our results are therefore compatible with a null evolution of the offsets with redshift between z=5-7, at fixed mass.

\subsection{Dust-UV offsets as a function of stellar mass}

\begin{figure}[!t]
\centering
\includegraphics[width=0.5\textwidth]{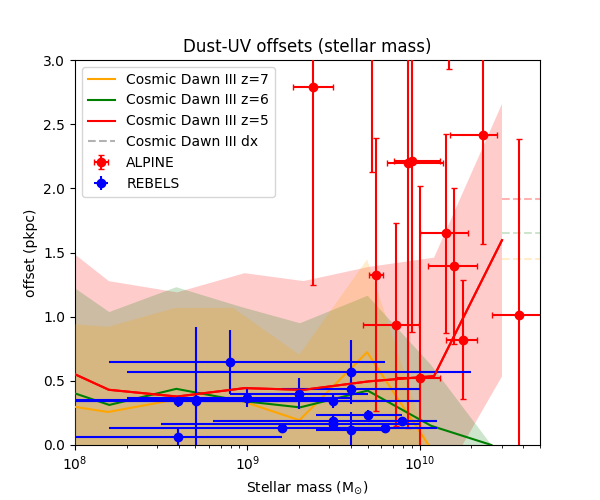}
\caption{Same as Fig. \ref{fig:dust_uv_offsets_mag} but as a function of stellar mass.}
\label{fig:dust_uv_offsets_stellarmass}
\end{figure}

We now present the dust-UV offsets as a function of the observed stellar masses. Figure \ref{fig:dust_uv_offsets_stellarmass} presents the relationship between dust-UV offsets and stellar mass for both our simulations and observational data. Several key features appear, predominantly in line with the previous sub-sections, but with some nuances however:

1. Trend with stellar mass: The dust-UV offsets become significantly large on average only for the latest epoch and the most massive galaxies of the simulation. This is especially visible at z=5.

2. Interestingly, there is a significant difference in this parameter space between the 2 observational samples, as the REBELS galaxies are on average less massive and display smaller offsets than their ALPINE counterparts. Strikingly, this difference is well described by CoDaIII galaxies. Indeed, CoDaIII galaxies at M$_*<10^{10}$ M$_\odot$ display small offsets (i.e. smaller than the simulation's resolution), compatible with a 0 average. The offset increases above this mass, and connects with the ALPINE data. Our results suggest that stellar mass is the main parameter driving the larger offsets of ALPINE as compared with REBELS.




3. Redshift evolution: as in the previous figures, there is no clear trend with redshift, beyond the fact that the latest epochs feature naturally more massive haloes. %
However, at fixed mass, the curves for each redshift fall well within  1 $\sigma$ of each other. Our results are therefore compatible with a null evolution of the offsets with redshift between z=5-7, at fixed stellar mass.

\section{Conclusions}
Our study of dust-UV offsets in high-redshift galaxies using the CoDa III simulation has yielded several important findings:

\begin{enumerate}

    \item{Reproducing the observed UV LF at $z = 5-7$ requires us to apply a reduction factor K$_{\rm dust}$=0.075 to all dust masses in CoDa III. Without this factor our UV LFs are too faint. With this reduction, we also find a fair agreement with the ALPINE and REBELS UV slopes $\beta$. 
    }
    \item Within this framework, the simulation successfully reproduces the observed spatial offsets between dust and UV emission in massive galaxies ($M_\mathrm{DM} > 10^{11.5}$M$_\odot$), in particular at z=5, with offsets increasing with dark matter halo mass and UV brightness.
    
    \item These offsets primarily result from severe dust extinction in galactic centers rather than a misalignment between dust and stellar distributions. The dust generally remains well-aligned with the bulk stellar component, and we predict comparatively small dust-NIR offsets.
    
    \item The amplitude of the offsets and their dependence on galaxy properties are in fair agreement with observational data from the ALPINE surveys, particularly for the brightest galaxies ($M_\mathrm{AB1600} < -21.5$) and at z=5.

    \item For these galaxies, the offsets we predict are comparable to or larger than the cell size and are, therefore, resolved by the CoDa III spatial grid.

    \item Our results suggest that the larger offsets of the ALPINE galaxies are mainly due to their larger stellar masses as compared to the REBELS sample.
    
    \item Our results also suggest that dust-UV offsets should be smaller for fainter galaxies $M_\mathrm{AB1600} > -21$), a prediction that can be tested by future deeper surveys and numerical simulations.
    
\end{enumerate}

These findings underscore the crucial role of dust in shaping the observed properties of high-redshift galaxies. They demonstrate that the complex dust-star geometry in early galaxies can lead to significant disparities between their apparent and intrinsic structures. This work provides a theoretical framework for interpreting current and future observations of galaxies during the epoch of reionization, and emphasizes the importance of considering dust effects when studying galaxy formation and evolution in the early universe.

Future studies should improve on spatial resolution beyond CoDaIII's, such as in \cite{Trebitsch2021} and \cite{thesanzoom}, as well as on the physical modelling of dust, as in \cite{dubois_dust}. Balancing high-enough spatial resolution inside of galaxies (e.g. down giant molecular clouds for instance) and large enough volumes to capture the largest galaxies remains a very ambitious, until now impractical goal. However, some hope is allowed thanks to  the increasing power of the new generation of exascale supercomputers. The future lies with a new generation of exascale astrophysical simulation codes, such as \cite{dyablo2024}, to further our understanding of the formation and evolution of galaxies during the epoch of reionization and along cosmic time.

\begin{acknowledgements}
We thank the anonymous referee for a thorough review and eye-opening comments which significantly improved the paper. This work used resources of the Oak Ridge Leadership Computing Facility, in particular the Summit supercomputer, under project AST031. An award of computer time was provided by the Innovative and Novel Computational Impact on Theory and Experiment (INCITE) program.
Our analysis made use of Python Jupyter notebooks, NumPy \citep{numpy}, Astropy \citep{astropy}, and PyVO. We also utilized the TAP VizieR service at CDS, querying the catalog 'J/A+A/643/A2' \citep{Bethermin2020}. K.A. is supported by NRF-2021R1A2C1095136 and RS-2022-00197685.
\end{acknowledgements}

\bibliographystyle{aa}
\bibliography{references}

\end{document}